\documentclass[longauth]{aa} 
\usepackage{subfigure}
\usepackage{graphicx}
\usepackage{epstopdf}
\usepackage{txfonts}
\usepackage{natbib}
\bibpunct{(}{)}{;}{a}{}{,} 

\def\prod{$\rm s^{-1}$}

\begin{document}
   \titlerunning{67P Churyumov-Gerasimenko}

   \title{Observations and analysis of a curved jet in the coma of comet 67P/Churyumov-Gerasimenko}

   \authorrunning{Zhong-Yi Lin\inst{1}, Wing-Huen Ip\inst{1,2,3}}
   \author{Z.-Y. Lin\inst{\ref{inst1}}
   \and I.-L. Lai\inst{\ref{inst2}}
   \and C.-C. Su\inst{\ref{inst3}}
   \and W.-H. Ip\inst{\ref{inst1},\ref{inst2},\ref{inst4}}
   \and J.-C. Lee\inst{\ref{inst5}}
   \and J.-S., Wu\inst{\ref{inst3}}
   \and J.-B. Vincent\inst{\ref{inst6}}
   \and F. La Forgia\inst{\ref{inst7}}
   \and H. Sierks\inst{\ref{inst6}}
   \and C. Barbieri\inst{\ref{inst7},\ref{inst8}}
   \and P. L. Lamy\inst{\ref{inst9}}
   \and R. Rodrigo\inst{\ref{inst10},\ref{inst11}}
   \and D. Koschny\inst{\ref{inst12}}
   \and H. Rickman\inst{\ref{inst13}}
   \and H. U. Keller\inst{\ref{inst14}}
   \and J. Agarwal\inst{\ref{inst6}}
   \and M. F. A'Hearn\inst{\ref{inst15}}
   \and M. A. Barucci\inst{\ref{inst16}}
   \and J.--L. Bertaux\inst{\ref{inst17}}
   \and I. Bertini\inst{\ref{inst8}}
   \and D. Bodewits\inst{\ref{inst15}}
   \and G. Cremonese\inst{\ref{inst18}}
   \and V. Da Deppo\inst{\ref{inst19}}
   \and B. Davidsson\inst{\ref{inst20}}
   \and S. Debei\inst{\ref{inst21}}
   \and M. De Cecco\inst{\ref{inst22}}
   \and S. Fornasier\inst{\ref{inst17}}
   \and M. Fulle\inst{\ref{inst23}}
   \and O. Groussin\inst{\ref{inst9}}
   \and P. J. Guti\'errez\inst{\ref{inst24}}
   \and C. G\"uttler\inst{\ref{inst6}}
   \and S. F. Hviid\inst{\ref{inst25}}
   \and L. Jorda\inst{\ref{inst9}}
   \and J. Knollenberg\inst{\ref{inst25}}
   \and G. Kovacs\inst{\ref{inst6}}
   \and J.-R. Kramm\inst{\ref{inst6}}
   \and E. K\"uhrt\inst{\ref{inst25}}
   \and M. K\"uppers\inst{\ref{inst10}}
   \and L.M. Lara\inst{\ref{inst24}}
   \and M. Lazzarin\inst{\ref{inst7}}
   \and J. J. L\'opez-Moreno\inst{\ref{inst24}}
   \and S. Lowry\inst{\ref{inst26}}
   \and F. Marzari\inst{\ref{inst7}}
   \and H. Michalik\inst{\ref{inst15}}
   \and S. Mottola\inst{\ref{inst25}}
   \and G. Naletto\inst{\ref{inst8},\ref{inst19},\ref{inst27}}
   \and N. Oklay\inst{\ref{inst6}}
   \and M. Pajola\inst{\ref{inst8}}
   \and A. Ro\.{z}ek\inst{\ref{inst8}}
   \and N. Thomas\inst{\ref{inst28}}
   \and C. Tubiana\inst{\ref{inst6}}
}

\institute{
   Institute of Astronomy, National Central University, Chung-Li 32054, Taiwan \email{zylin@astro.ncu.edu.tw} \label{inst1}
   \and Institute of Space Sciences, National Central University,  Chung-Li 32054, Taiwan \label{inst2}
   \and Department of Mechanical Engineering, National Chiao Tung University, Taiwan \label{inst3}
   \and Space Science Institute, Macau University of Science and Technology, Macau \label{inst4}
   \and Dept. of Earth Science, National Central University, Chung-Li 32054, Taiwan \label{inst5}
   \and Max-Planck Institut f\"ur Sonnensystemforschung, Justus-von-Liebig-Weg, 3, 37077 G\"{o}ttingen, Germany \label{inst6}
   \and Department of Physics and Astronomy "G. Galilei", University of Padova, Vic. Osservatorio 3, 35122 Padova, Italy \label{inst7}
   \and Centro di Ateneo di Studi ed Attivit\'{a} Spaziali "Giuseppe Colombo" , University of Padova, Via Venezia 15, 35131 Padova, Italy \label{inst8}
   \and Aix Marseille Universit\'e, CNRS, LAM (Laboratoire dAstro-physique de Marseille)UMR 7326, 13388, Marseille, France \label{inst9}
   \and Centro de Astrobiologia (INTA-CSIC), European Space Agency (ESA), European Space Astronomy Centre (ESAC), P.O. Box 78, E-28691 Villanueva de la Ca\~nada, Madrid, Spain \label{inst10}
   \and International Space Science Institute, Hallerstrasse 6, 3012 Bern, Switzerland \label{inst11}
   \and Research and Scientific Support Department, European Space Agency, 2201 Noordwijk,The Netherlands  \label{inst12}
   \and PAS Space Reserch Center, Bartycka 18A, 00716 Warszawa, Poland \label{inst13}
   \and Institut f\"{u}r Geophysik und extraterrestrische Physik (IGEP), Technische Universit\"{a}t Braunschweig, 38106 Braunschweig, Germany \label{inst14}
   \and Department for Astronomy, University of Maryland, College Park, MD 20742-2421, USA \label{inst15}
   \and LESIA-Observatoire de Paris, CNRS, UPMC Univ Paris 06, Univ. Paris-Diderot, 5 Place J. Janssen,  92195 Meudon Pricipal Cedex, France \label{inst16}
   \and LATMOS, CNRS/UVSQ/IPSL, 11 Boulevard dAlembert, 78280 Guyancourt, France  \label{inst17}
   \and INAF Osservatorio Astronomico di Padova, vic. dell?Osservatorio 5, 35122 Padova, Italy  \label{inst18}
   \and CNR-IFN UOS Padova LUXOR, via Trasea 7, 35131 Padova, Italy \label{inst19}
   \and Department of Physics and Astronomy, Uppsala University, Box 516, 75120 Uppsala, Sweden \label{inst20}
   \and Department of Industrial Engineering, University of Padova, Via Venezia 1, 35131 Padova, Italy \label{inst21}
   \and University of Trento, via Sommarive 9, 38123 Trento, Italy \label{inst22}
   \and INAF -- Osservatorio Astronomico di Trieste, via Tiepolo 11, 34143 Trieste, Italy \label{inst23}
   \and Instituto de Astrof\'{\i}sica de Andaluc\'{\i}a (CSIC), c/Glorieta de la Astronom\'{\i}a, 18008 Granada, Spain \label{inst24}
   \and Deutsches Zentrum f\"{u}r Luft- und Raumfahrt (DLR), Institut f\"{u}r Planetenforschung, Rutherfordstrasse 2, 12489 Berlin, Germany \label{inst25}
   \and Centre for Astrophysics and Planetary Science, School of Physical Sciences, The University of Kent, Canterbury CT2 7NH, United Kingdom\label{inst26}
   \and Department of Information Engineering, University of Padova, via Gradenigo 6/B, 35131 Padova, Italy \label{inst27}
    \and Physikalisches Institut der Universit\"{a}t Bern, Sidlerstr. 5, 3012 Bern, Switzerland  \label{inst28}
}
   \date{Received:...; accepted: ... }

  \abstract
   {} 
   {Analysis of the physical properties and dynamical origin of a curved jet of comet 67P/Churyumov-Gerasimenko that was observed repeatedly in several nucleus rotations starting on May 30 and persisting until the early-August, 2015. }
   {Simulation of the motion of dust grains ejected from the nucleus surface under the influence of the gravity and viscous drag effect of the expanding gas flow from the rotating nucleus.}
    {The formation of the curved jet is a combination of the size of the dust particles ($\sim$0.1-1 mm) and the location of the source region near the nucleus equator, hence enhancing the spiral feature of the collimated dust stream after being accelerated to a terminal speed of the order of m s$^{-1}$.\\
}
  {}
%

\maketitle

\section{Introduction}\label{intro}
Comets, as the most primitive bodies from the outer region of the solar system, are known to be very abundant in volatile ices and refractory dust grains. For a comet with orbital parameters inside 3-4 au, solar radiation raises the surface temperature to such an extent that ice sublimation initiates, as indicated by the appearance of a coma containing expanding gas and small dust particles. Therefore, the outgassing activity increases as the comet approaches perihelion. The behavior of comet 67P Churymov-Gerasimenko (67P hereafter) follows this pattern closely (\cite{2011A&A...525A..36L}; \cite{2011A&A...531A..54T}; \cite{2013A&A...549A.121V}). In addition to the nearly spherically symmetric coma, anisotropic structures in the form of collimated jets have been identified (\cite{2011A&A...525A..36L}; \cite{2013A&A...549A.121V}). The identification of the source mechanism and acceleration process of these dust jets are one of the main focus of the Rosetta mission.

The close-up observations of the OSIRIS scientific camera on the Rosetta spacecraft (\cite{2007SSRv..128..433K}) have provided an unprecedented view of the morphology of the near-nucleus coma (\cite{Sierks2015}; \cite{Thomas2015}). The jet feature were visible in the early phases of the rendezvous mission (\cite{2015A&A...583A..A11}; \cite{2015A&A...583A..A9}), and their development was closely monitored since then. Due to the orientation of its rotational axis and its complex shape, and a rotation period of 12.4 hours (Mottola et al. 2015), significant diurnal variations and seasonal effect of the gas flow and dust coma structure were detected (\cite{Gulkis2015};  \cite{2015Sci...347a0276H}). 

\begin{figure}
\includegraphics[width=0.5\textwidth]{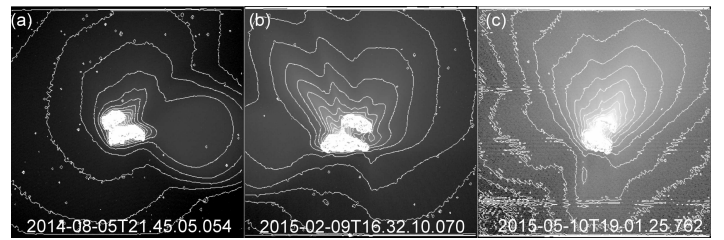}
\caption{Gradual growth of the dust coma and dust jets of comet 67P: (a) image taken on 5 August, 2014, when the heliocentric distance r$_h$= 3.60 au, only a small dust jet can be seen emanating from the Hapi region; (b) 9 February, 2015 for r$_h$ = 2.35 au, the formation of a main jet is accompanied by a few fainter jets originating from other regions; (c) 10 May, 2015 for r$_h$ = 1.67 au, a system of bright jets appearing on the sunward side of the coma. The image contrast level is adjusted to log scale, ranging from -4 to -7.}
\label{JetsTime}
\end{figure}

Figure 1 shows the time development of the dust coma and jets as 67P approaches the Sun. From August 2014 ($r_h$ $\sim$3.60 au) to May 2015 ($r_h$ $\sim$1.67 au), the dust coma became more dense with the same image contrast level. At the beginning, the Hapi region located in the neck between the two lobes appeared to be the main source of the water gas flow (\cite{Gulkis2015}; \cite{2015A&A...583A..A3}; \cite{2015Sci...347a0276H}) and dust jets (\cite{Sierks2015}; \cite{2015A&A...583A..A11}; \cite{2015A&A...583A..A9}; \cite{Vincent2015a}). All the dust jets have very straight configurations perpendicular to the surface, suggesting efficient acceleration of the embedded solid grains to radial speeds far exceeding the angular velocity ( $V_r$$\sim$1-2 m s$^{-1}$) of the nucleus due to its rotation. It was therefore surprising that a jet structure with large curvature appeared in late May (Fig. 2). This is the first time that a spiral structure was seen in the near-coma region of a comet made possible because of the close distance of the Rosetta spacecraft to the comet nucleus. Nevertheless, from ground-based observations (\cite{1991Icar...93..194S}, \cite{2007A&A...469..771L}, \cite{2012A&A...537A.101L}, \cite{2013AJ....146....4L}), we did see the repeatability of a curved appearance of the gaseous and dust jets related to the rotation of the nucleus or of the dust jets forming to dust tail due to the solar gravity and radiation pressure. However, these ground-based observations have much larger scales than Rosetta observations and the important physical processes might not be the same as the curved jets observed by ROSETTA. The curved jets persisted about two months and disappeared in early-August, 2015. In this work, we will examine the observed properties and dynamics of the curved jet as well as the localization of its possible source region. 

The paper is organized as follows. Section 2 describes the morphology and time variation of the curved jet in comparison to other collimated dust jet features. The results of a set of computations making use of the gravity field model of the comet nucleus and of the Direct Simulation Monte Carlo (DSMC) simulation are presented in Section 3. A discussion on the theoretical results and the OSIRIS imaging data is given in Section 4.

\section{Observations and data analysis}\label{obs}
\subsection{Observations on May 30-31}
Figure 2 shows the time sequence of the dust coma of 67P in approximately one nucleus rotation on May 30-31, 2015. The sun is towards the top and that is the reason why a clear shadow was cast behind the nucleus. The spin axis was pointing away from the projection plane with the rotation in the clockwise direction. According to the shape model (\cite{Sierks2015}; \cite{2015A&A...583A..A33}; \cite{Jorda2015}), the sub-solar point during this observation was close to the equator of comet 67P. We note that three different components can be identified in Fig.2.a. The brightest one (J1) can be traced to the middle of the Hapi region. On its left side an array of jets (e.g., J2) became clearly visible in Fig. 2.b because of the sun-lit effect on Imhotep. Last but not least, on the right-hand side a jet (J3) in the form of a spiral arm came into view in Fig. 2.b. It is less discernable in subsequent images as a result of the orientation of the spacecraft relative to the rotating nucleus and/or decrease of the dust production.  The spiral jet reappeared again in Fig. 2.h just one rotation period later.

\begin{figure}
\begin{center}
\includegraphics[width=0.5\textwidth]{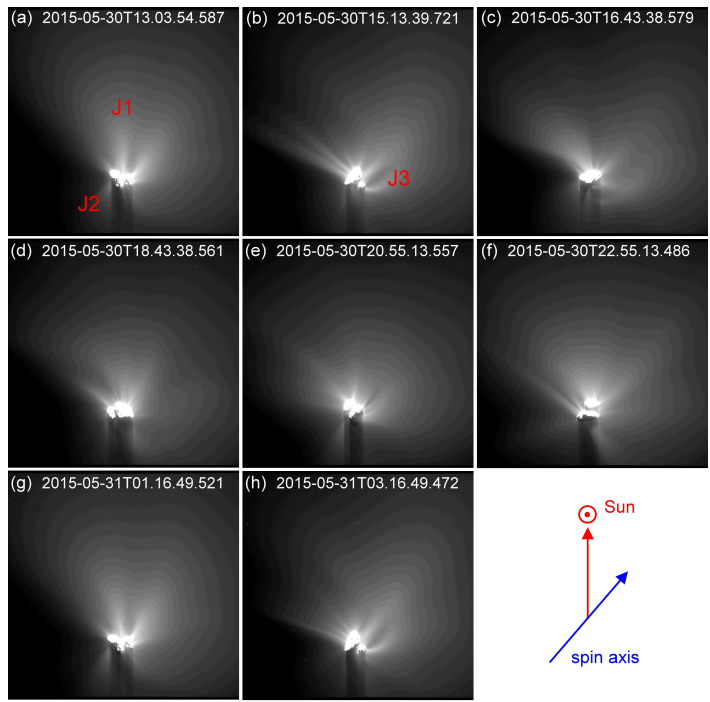}
\end{center}
\caption{Jet structures obtained with the wide-angle camera from 13:03 UT on May 30 to 07:28 UT on May 31, 2015. Sub-panels (a) to (h) are separated by about two hours between two frames. The spatial scales and field of view range from 6.72 m/px,13.75 km (fig. a) to 7.01 m/px, 14.36 km (fig. h). }
\label{CurvedJetAll}
\end{figure}

\subsection{Source region of the Curved jet}
In order to find the source region of the curved jet, we both used the method described in Lin et al (2015) and the jet inversion method tracing the orientation of the colliminated beams back to their emission points (\cite{Vincent2015b}, \cite{2015A&A...583A..A9}). The shape model constructed by Jorda et al (2015) and current SPICE kernels have been used for this purpose. Nonetheless, it is difficult to obtain a precise location of the source region from the OSIRIS images taken from late-May to early-June, 2015 because of the diffusive structure of the jet. The most probable source region is located between Nut and Serget (see Figure 3 for a context image of where these regions are located on the nucleus). At closer scrutiny, it can be seen that this region is covered by smooth deposits of fine materials which are likely the result of airfall of low velocity particles, not being able to escape from the nucleus surface (\cite{Thomas2015}, \cite{2015A&A...583A..A41} and reference in). However when looking closely at Nut region there is a more granular deposit including boulders with diameters up to a few tens of meters. A part of this area contains "pit-like" features with sub-meter to 4-5 meter diameter (see fig. 7 in \cite{2015A&A...583A..A41}) that might be remnants formed by wind erosion or sublimation of the volatile-rich blocks. The curved jet appeared for two more months since the first detection. As a consequence, the features of the source region surface might have changed. A study of possible changes in localized areas of the identified source region would however be possible only in the later phase of the Rosetta mission, when high resolution images at spatial scales as small as 0.5 m/px will be available again.

\begin{figure}
\begin{center}
\includegraphics[width=0.5\textwidth]{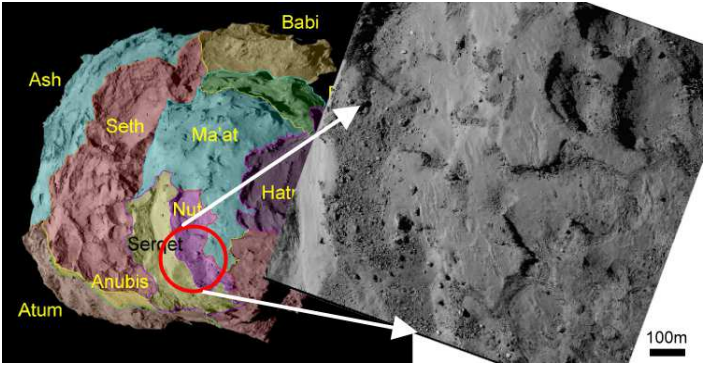}
\end{center}
\caption{Geological map of 67P observed from the top of the head lobe and nomenclature of the geological regions (modified from Fig. 2 in \cite{2015A&A...583A..A26}  and the possible source regions, red-circles, of the curved jet. The latitude of the possible source region is about 6 degree at the northern hemisphere. The subsolar point at the time of the present observation (31, May, 2015) is 9 degree at the southern hemisphere and moves toward to higher latitude until the beginning of September, 2015. The right-panel is the NAC image obtained on September 19.5, 2014 with a resolution about 0.53 m/px.}
\label{SR_CurvedJet}
\end{figure}

\section{Numerical Simulation}
To examine the origin and dynamical evolution of the curved jet, we need to consider the trajectories of different dust particles with different sizes under the influence of the gravitational attraction of the nucleus. Because of its highly irregular shape, the gravitational field have been computed by dividing the whole object into 33681 elements - according to the shape model with homogeneous structure and a bulk density of 532 kg m$^{-3}$ (\cite{Jorda2015}). Figure 4.a shows the contour plot of the gravitational field in the vicinity of the nucleus. At distances larger than three nucleus radii ($\sim$6 km) the gravitational field can be reasonably approximated by that from a point mass. However, close to the nucleus surface, the field distribution is far from spherical symmetry. 

Figure 4.b shows the flow field of the coma gas expanding from the nucleus surface. To treat the transition from a collisional region close to the central nucleus to the collisionless coma at large distance, the Direct Simulation Monte Carlo (DSMC) method is required (\cite{Bird1994}). The basic structure of the DSMC code used for obtaining this result has been described in detail in \cite{Wu2004166}, \cite{Su20101136}, \cite{Lai2016} et al,  and \cite{Liao2015} and will not be repeated here. For this simulation, we assume the water production rate is 10$^{27}$ molecules s$^{-1}$ and the sunlit portion of the nucleus surface at the time of consideration has been assumed to be all active in outgassing. Thus, a uniform gas production rate in sunlit side is Z = 4.8$\times$10$^{19}$ H$_2$O molecules m$^{-2}$ s$^{-1}$. The initial velocity distribution of the gas is described by a half-Maxwellian distribution with a thermal temperature of 228 K. The sunlit side is assumed to be free of gas outflow even though we know this is not necessarily true according to both  the Rosina measurements (\cite{2015Sci...347a0276H}) and the recent discovery of the so-called night-side outbursts or Sunset jet activities by the OSIRIS camera team (\cite{Knollenberg2015} submitted and \cite{Xian2016}). It is clear that a certain level of weak outgassing activity existed on the nightside of the nucleus.  Also, there can be activity driven by other gases (CO$_2$) - as may be the case for both the sunrise jets and for the sunset jets, and as was seen by Haessig (CO$_2$/H$_2$O = 4). However, even under such circumstances, the number distribution of the gas molecules should be highly non-isotropic within 5-10 nucleus radii covered by the simulation box. In a different study (\cite{Lai2016} et al), it has been demonstrated that the global gas flows tend to follow streamlines perpendicular to the surface of their source regions if the gas emission rate (Z) on the illuminated side is assumed to be proportional to square root $\cos \theta$ where $\theta$ is the solar zenith angle where $\theta$ $>$0.  For the whole surface, Z = 0.1$\times$ Z$_0$ where Z$_0$ is the peak sublimation rate at $\theta$= 0. It is interesting to note that, even for localized outburst events, the dust jets appeared to be highly collimated (J-B.Vincent, private communication, 2016).

\begin{figure}
\begin{center}
\includegraphics[width=0.4\textwidth]{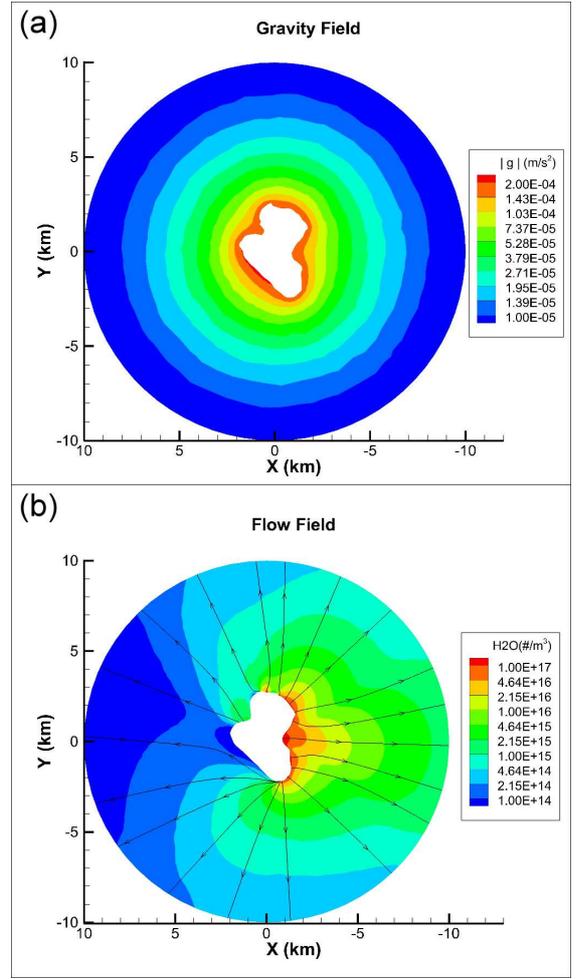}
\end{center}
\caption{ (a) A contour plot of the gravitational field of comet 67P in the XY plane containing the long and short axes. The rotational axis is in the perpendicular direction. The assumed bulk density is $\rho$= 532 kg m$^{-3}$. (b) The flow field of the expanding gas outflow. }
\label{CurvedJet_GG}
\end{figure}

In the simulation code, the motion of a dust particle of mass m with initial zero velocity is determined by the viscous drag effect of the expanding gas flow and the gravitational attraction of the nucleus. Note that the effects of solar radiation pressure on curved jet are not considerable in our simulation but will be involved in the future. The used equation of motion is shown below. (\cite{Gombosi1986}; \cite{1999Icar..140..173S}; \cite{2008EM&P..102..521M}). \\
\begin{equation}\label{color_measure}
m\frac{dv}{dt} = m\vec{g}+\frac{1}{2}A\rho_gC_dv_r^2
\end{equation}

where m is the mass of dust particle, and g is the "local" gravity. The second term on the right-hand side represents the gas drag effect. The dust particle of cross section A is assumed to be spherical and its density ($\rho$) is taken to be 1000 kg m$^{-3}$ (\cite{2015ApJ...802L..12F}). In Equation 1, v$_r$ is the relative velocity between the gas molecules and the dust particle, $\rho$$_g$ is the mass density of gas flow, and C$_d$ is the drag coefficient. Note that C$_d$ = 2 (\cite{Wallis1982}).  

To demonstrate the combined effect of the gas drag and nucleus gravity, a source region is tentatively chosen in the Nut region (see Figure 3) with zero initial velocity. The general idea is to examine the dependence of particle size, gas production rate and emission location on the jet dynamics. 

\section{Result and conclusions}
Figure 5.a illustrates the obtained velocity profiles of dust grains with radius ranging between 1 $\mu$m and 1 mm for the gas sublimation rate of Z = 4.8$\times$10$^{19}$ H$_2$O molecules m$^{-2}$ s$^{-1}$. The bigger dust grains will be accelerated to lower radial velocity as compared with the particles of smaller sizes. This effect can be understood in terms of the size dependence of the gas drag force under the influence of the gravitational attraction of the comet nucleus. In addition, the dust grains are accelerated to their terminal speeds within a distance of the order of 2 km or slightly more in all the considered cases. The small micron-sized dust could reach an outflow speed as high as 30 m s$^{-1}$ while those of mm-size have terminal speed of the order of 0.5 m s$^{-1}$ (or less) which is comparable to the rotational speed of the nucleus at its surface. This immediately suggests that the grains in the curved jet must be relatively large (i.e, d $\sim$0.1-1 mm). The result is consistent with previous findings (\cite{2015Sci...347a3905R}) that the optically dominant particles in 67P coma are exactly those of size of 0.1-1 mm. In additional to the curved jet, the straight jets (Fig. 2 J1 and J2) might consist of relative smaller particles. The identification of the exact sizes would depend on the characteristics of the surface material and the effective outgassing rate which controls the gas drag acceleration. 

\begin{figure}
\begin{center}
\includegraphics[width=0.4\textwidth]{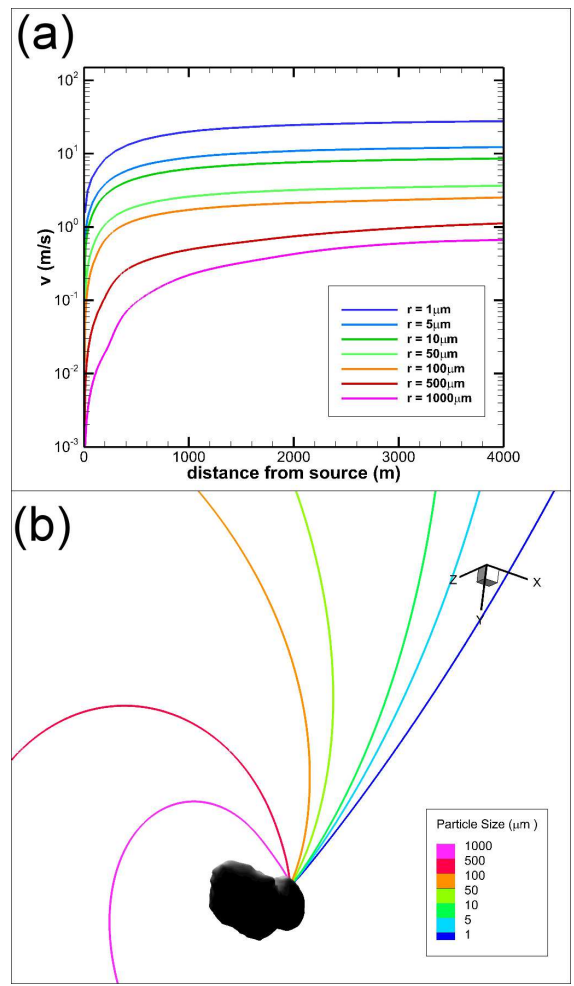}
\end{center}
\caption{(a)The velocity profiles of dust grains of different sizes with sublimation rate Z = 4.8$\times$10$^{19}$ H$_2$O molecules m$^{-2}$ s$^{-1}$. (b) Trajectories of dust grains of different radii (from 1$\mu$m to 1mm) as the nucleus rotates. }
\label{CurvedJet_V}
\end{figure}

It is perhaps not an accident that, the curved jet was observed to be emitted near the equatorial region of the head of 67P (Nut, Serqet and  Ma'at ), i.e. where the nucleus rotation speed is the largest. As mentioned before in Section 2, the collimated jets of linear configuration seemed to have been all emanated from the Hapi region that is more or less along the spin axis, where the centrifugal force is minimal.

From our data analysis and preliminary numerical simulation, it is now understood that the appearance of a curved jet in May and June 2015 is caused by a combination of the ejection of mm-sized dust grains from the equatorial source region in the vicinity of Nut, Serqet and Ma'at. This unique set of OSIRIS observations provides important information on the physical properties of the dust grains and on the acceleration process.

\begin{acknowledgements}
\tiny
OSIRIS was built by a consortium led by the Max-Planck-Institut f\"{u}r Sonnensystemforschung, G\"{o}ttingen, Germany, in collaboration with CISAS, University of Padova, Italy, the Laboratoire d'Astrophysique de Marseille, France, the Instituto de Astrof\'{\i}sica de Andaluc\'{\i}a, CSIC, Granada, Spain, the Scientific Support Office of the European Space Agency, Noordwijk, Netherlands, the Instituto Nacional de T\'{e}cnica Aeroespacial, Madrid, Spain, the Universidad Polit\'{e}chnica de Madrid, Spain, the Department of Physics and Astronomy of Uppsala University, Sweden, and the Institut f\"{u}r Datentechnik und Kommunikationsnetze der Technischen Universit\"{a}t Braunschweig, Germany. The support of the national funding agencies of Germany (Deutschen Zentrums f\"{u}r Luft- und Raumfahrt), France (Centre National d'Etudes Spatales), Italy (Agenzia Spaziale Italiana), Spain (Ministerio de Educaci\'{o}n, Cultura y Deporte), Sweden (Swedish National Space Board; grant no. 74/10:2), and the ESA Technical Directorate is gratefully acknowledged. This work was also supported by grant number NSC 102-2112-M-008-013-MY3 and NSC 101-2111-M-008-016 from the Ministry of Science and Technology of Taiwan.  We are indebted to the whole Rosetta mission team, Science Ground Segment, and Rosetta Mission Operation Control for their hard work making this mission possible.
\end{acknowledgements}

\bibliographystyle{aa} 
\bibliography{OSIRIS_CurvedJet}    

\end{document}